\documentclass[]{spie}  

\newcommand*\arcsec{\ensuremath{^{\prime\prime}}}

\usepackage{amsmath,amsfonts,amssymb}
\usepackage{graphicx}
\usepackage[colorlinks=true, allcolors=blue]{hyperref}

\title{An interferometric view of binary stars}

\author[a]{Henri M.J. Boffin}
\affil[a]{ESO, Karl-Schwarzschild-str. 2, 85748 Garching, Germany}

\authorinfo{E-mail: hboffin@eso.org}

\begin{document} 
\maketitle

\begin{abstract}
The study of binary stars is critical to apprehend many of the most interesting classes of stars. Moreover, quite often, the study of stars in binary systems is our only mean to constrain stellar properties, such as masses and radii. Unfortunately, a great fraction of the most interesting binaries are so compact that they can only be apprehended by high-resolution techniques, mostly by interferometry. I present some results highlighting the use of interferometry in the study of binary stars, from finding companions and deriving orbits, determining the mass and radius of stars, to studying mass transfer in symbiotic stars, and tackling luminous blue variables. In particular, I show how interferometric studies using the PIONIER instrument have allowed us to confirm a dichotomy within symbiotic stars, obtain masses of stars with a precision better than 1\%, and help us find a new $\eta$ Carinae-like system. I will also illustrate the benefits for the study of binary stars one would get from upgrading the VLT Interferometer so as to be able to observe in the visible range.  
\end{abstract}


\section{INTRODUCTION}
\label{sec:intro}  

Stars tend to form and live in binaries, with the fraction of binary stars being about 50\% for solar-like stars to more than 70\% for OB stars. A significant fraction of these binary stars will interact in one way or another during their evolution.  Mass transfer will affect the chemical composition of the companion, which can for example become polluted in processed elements like in Algols or in Barium stars. It also leads to a mass increase of the companion, with sometimes strong consequences, such as blue straggler stars or Type Ia supernovae. Finally, mass transfer has an impact on the evolution of the orbital separation. The study of binary stars is therefore critical to apprehend many of the most interesting classes of stars. Moreover, the study of stars in binary systems is often our only means of constraining stellar properties, such as masses and radii. 

A great fraction of the most interesting binaries are so compact that they can be apprehended by high-resolution techniques only, in particular by interferometry.
To understand this, one should realise that in order to have the most complete information of the components of a binary system, one need to combine a spectroscopic orbit with a visual one. In both cases, ideally, one needs to resolve both components. 

If limited by seeing, i.e. turbulence of the atmosphere from the ground, the spatial resolution that can be achieved is around 0.5$\arcsec$. For a star at 100 pc (typical of many low-mass stars or red giants, for example), this would correspond to separations of the order of 50 au, and thus orbital periods around 100 years. For a massive star, whose most interesting ones are often located at 1.5 kpc or more, this is even worse, with the typical separations being about 450 au and the orbital periods close to 2\,000 years. These are not very interacting binaries, i.e. not the most interesting ones. Using a space telescope or adaptive optics on the largest ground-based telescopes, one can now attain a factor 10 better in spatial resolution, which translates for a 100 pc system to separations of 5 au and periods of 10 years, which is already much better, but for massive stars, these numbers are still, respectively, 75 au and 100 years -- too large for comfort. So it is only by using interferometry and being able to reach the milli-arcsecond (mas) resolution, that one can now probe low-mass binaries at 100 pc with orbital periods of a few days, or massive stars binaries with separations of 1 au and periods of a few months. Interferometry is thus the key!  

 Below are some examples of how interferometry, and more particularly $H$-band observations with the {\nobreak PIONIER} instrument and the four 1.8 m Auxiliary Telescopes of ESO's Very Large Telescope Interferometer (VLTI), allowed us to get important results on different classes of binary stars.

 \begin{figure} [htbp]
   \begin{center}
   \begin{tabular}{ccc} 
   \includegraphics[width=5.3cm]{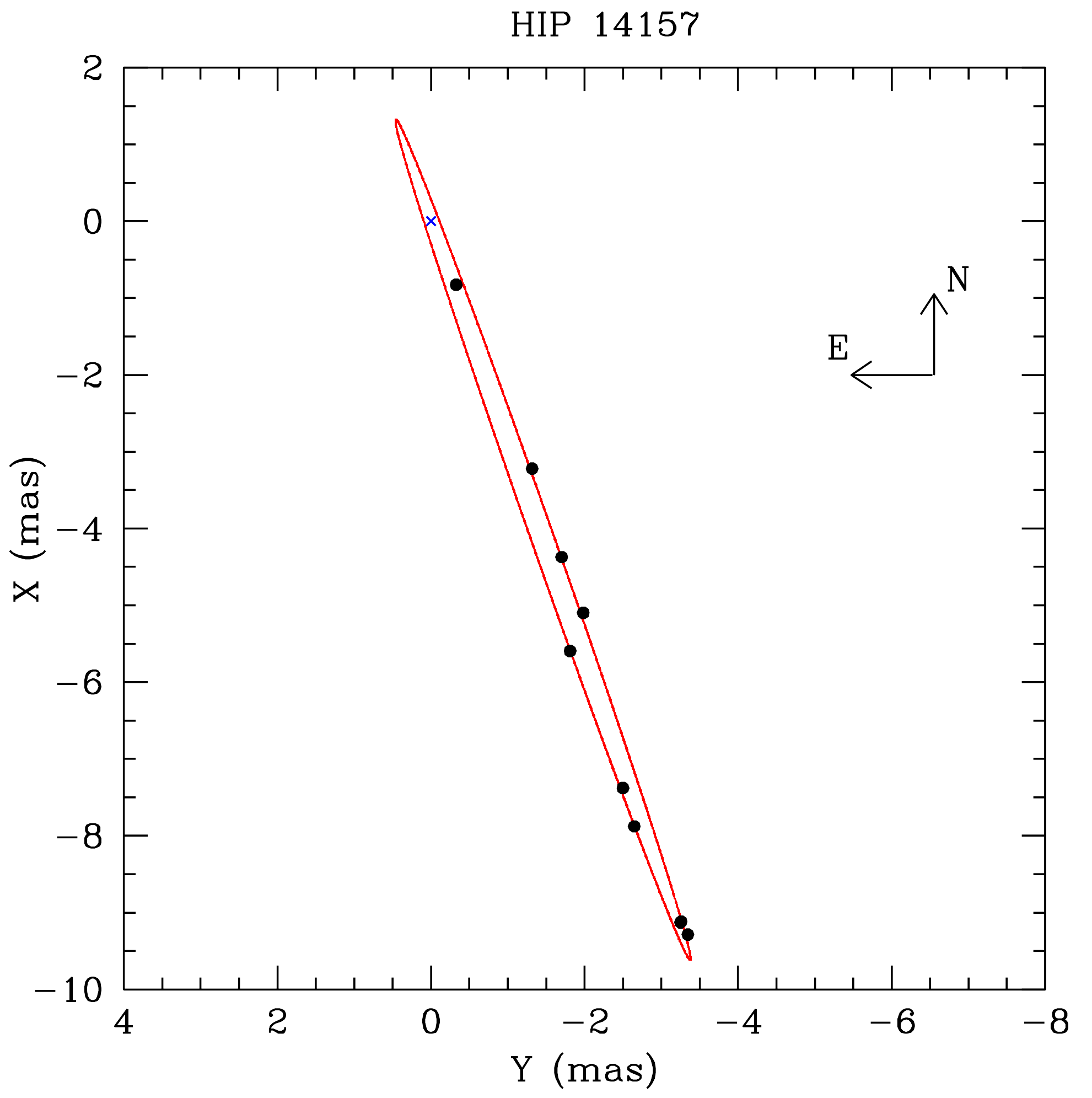}&\includegraphics[width=5cm]{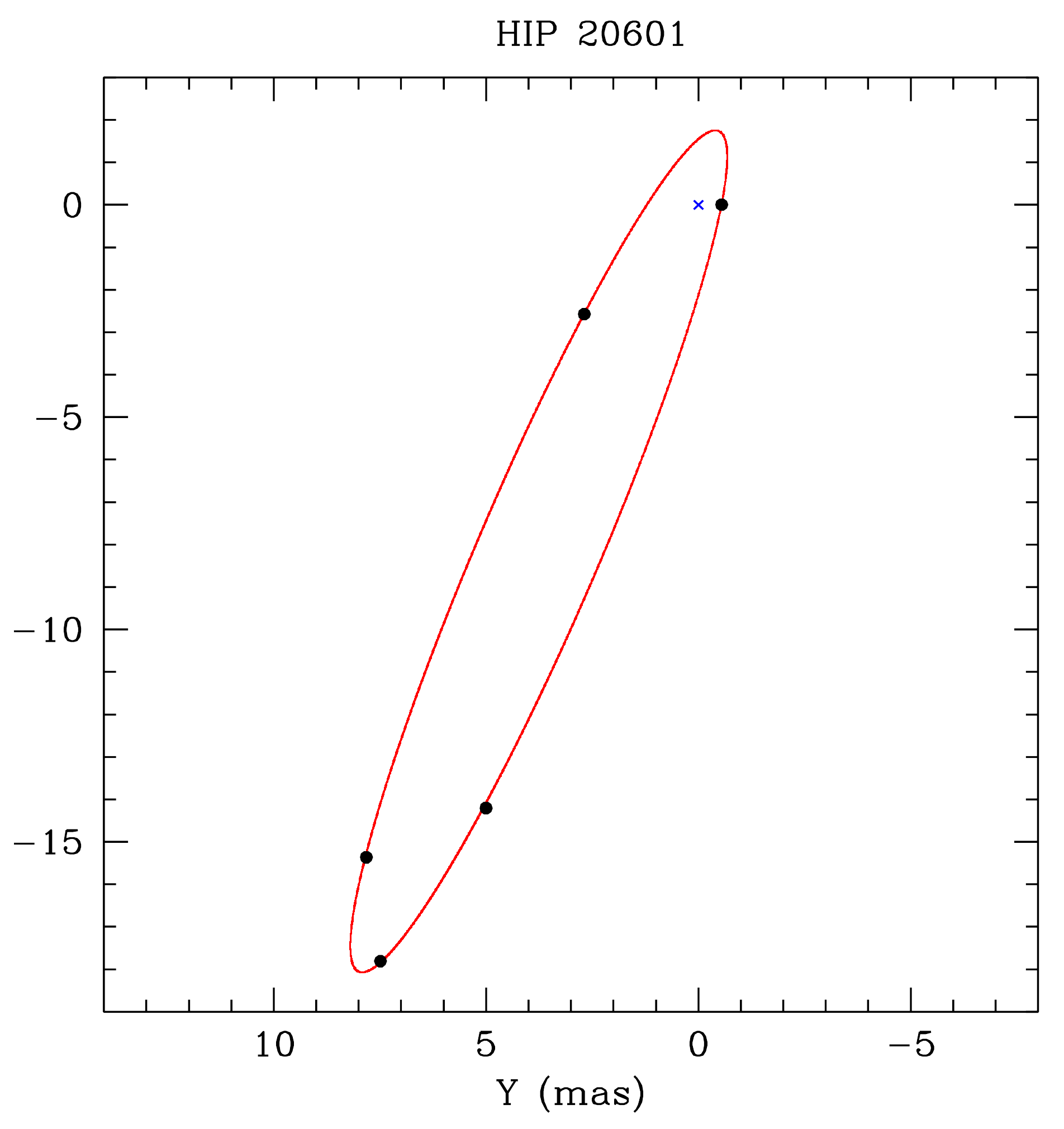}&\includegraphics[width=5cm]{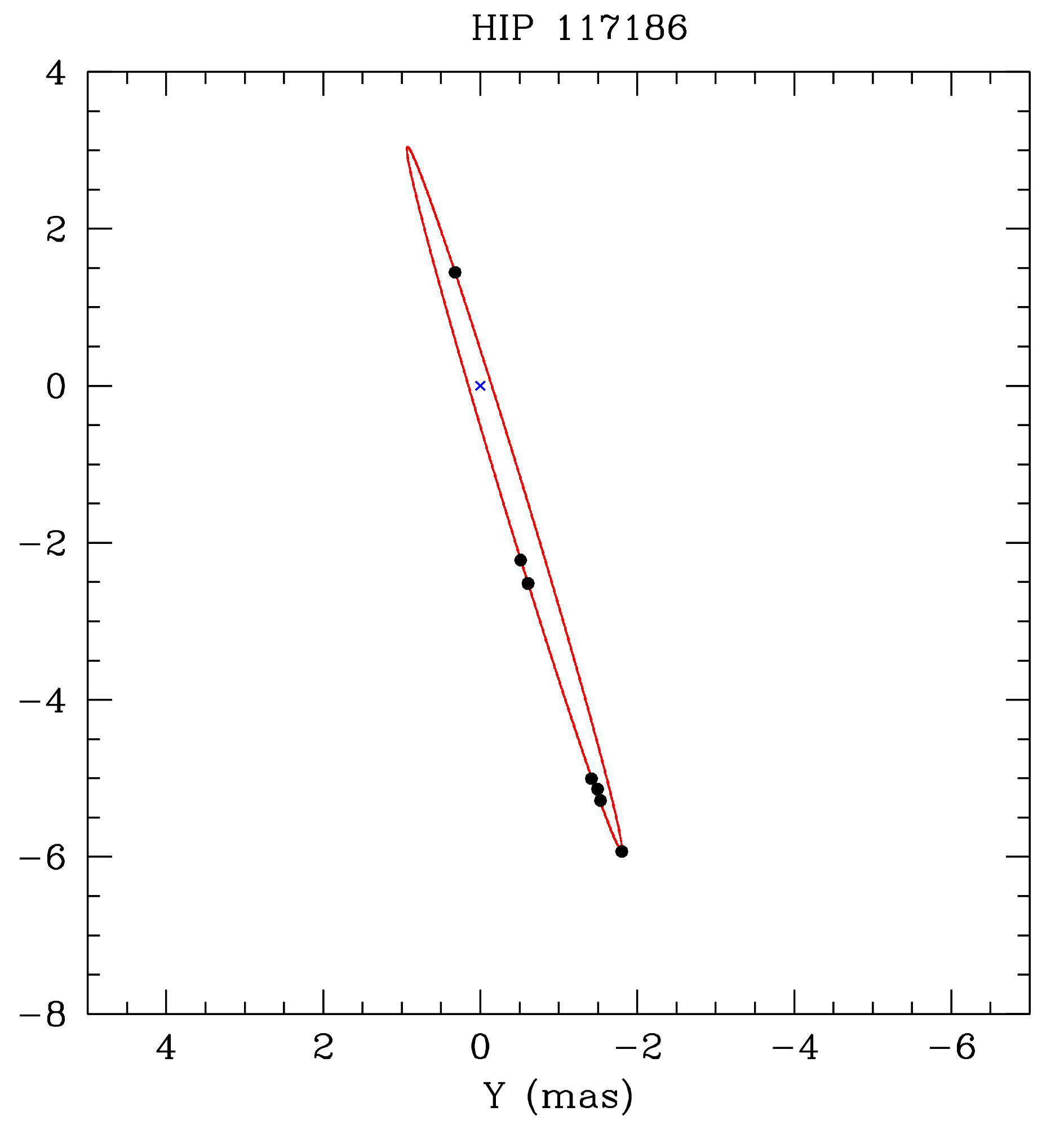}\\
   \end{tabular}
   \end{center}
   \caption
   { \label{fig:sb2} 
The visual orbits of the three SB2s observed with PIONIER. The position of the primary is indicated by the cross, while the observed positions correspond to the heavy dots. Adapted from Ref.~\citenum{Halbwachs16}.}
   \end{figure} 

\section{Precise stellar masses}
Halbwachs et al.\cite{Halbwachs14} presented the selection of a sample of double-lined spectroscopic binaries (SB2) for which they are obtaining precise radial velocities, with the final aim to derive at the 1\% level or better the masses of the components once the Gaia astrometric measurements will be available. However,  based on the reduction of the Hipparcos satellite, it is known that large space astrometric surveys may be prone to systematic errors.  Thus, an independent derivation of the masses of some stars from their sample is needed in order to validate all future results. This can be done thanks to interferometry. 

Using the PIONIER instrument, we are monitoring six SB2 to derive a visual orbit, which when combined with the spectroscopic radial velocities allow us to determine the individual masses of the components. A first set of results is presented in Refs. \citenum{Halbwachs15} and \citenum{Halbwachs16} for three binary systems, which we summarise here. 

Observations were done at 
several epochs (between 6 and 10, depending on the system) that covered between 70\% of the orbital period (for two systems) and more than twice the period for the last one. These interferometric observations were adjusted by a simple binary model, as the individual components, all smaller than 0.21 mas, are unresolved by the VLTI. The free parameters are thus the separation, the position angle of the secondary with respect to the primary and the flux ratio. We obtained average errors on the positions of about 0.01 mas for the best case, up to 0.07 mas for the worst one -- they are thus rather similar to the errors expected for Gaia. The flux ratio between the two components of each system was between 0.4 and 0.67, corresponding to a difference of 1 and 0.4 magnitudes, respectively.  

A minimum of eight observations is needed to derive all the parameters of a visual orbit and to estimate their errors. Since a relative position as obtained with interferometry is a two-dimensional observation, this corresponds to four interferometric observations, and we have generally much more than this.

Figure~\ref{fig:sb2} shows the observations as derived from the PIONIER data, as well as the final visual orbits derived. Combined with the spectroscopic orbits, this leads to a derivation of the masses of the components, as well as the parallax of the system, which are given in Table~\ref{Tab:mass}.

\begin{table}[htbp]
\caption{\label{Tab:mass} Derived masses and parallax for the three SB2 observed with PIONIER}
\begin{center}
\begin{tabular}{l|lll}
\hline
& HIP 14157 & HIP 20601 & HIP 117186 \\
\hline
M$_1$ (M$_\odot$)& 	0.982$\pm$0.010 	& 0.9808$\pm$0.0040 & 	1.686$\pm$0.021\\
M$_2$ (M$_\odot$)& 0.8819$\pm$0.0089 	& 0.7269$\pm$0.0019 	& 1.390$\pm$0.034\\
parallax (mas) 	& 19.557$\pm$0.078 	& 16.702$\pm$0.037 & 	8.445$\pm$0.075\\
\hline
\end{tabular}
\end{center}
\end{table}
	
The accuracy of the masses we obtained is between 0.26\% and 2.4\%, close to the uncertainties that we expect to obtain when combining radial velocity measurements with Gaia astrometry. Our sample will thus be a very good benchmark to test Gaia results. 

Quite noteworthy, five of the six derived masses are a few per cent larger than the expectations coming from the standard spectral type-mass calibration, while the parallaxes we obtain confirm those found in the Hipparcos 2 catalogue, although we have a   much better accuracy. Finally, our results indicate that HIP 14157 should be visible as an eclipsing binary and we urge observers to try to observe such eclipses, as this would confirm our inclination, improve the accuracy of the masses, and allow an estimation of the radii of the components. 

\begin{figure} [ht]
   \begin{center}
   \begin{tabular}{c} 
   \includegraphics[width=15cm]{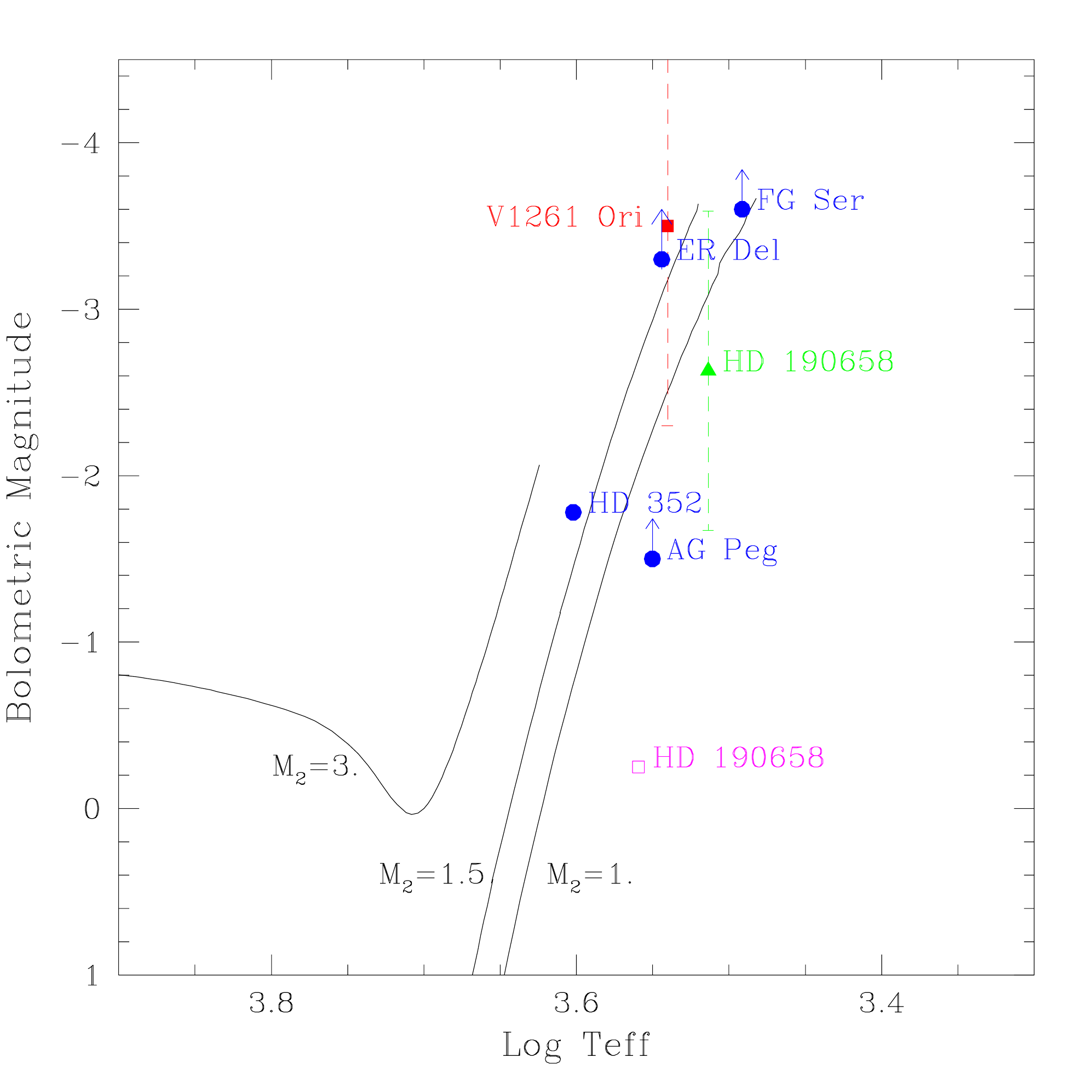}
   \end{tabular}
   \end{center}
   \caption 
   { \label{fig:diam} 
Hertzsprung-Russell diagram showing the positions of the 
symbiotic stars we observed with PIONIER (full dots, in 
blue), together with Yonsei-Yale evolutionary 
tracks for solar abundance stars with initial 
masses of 1, 1.5, and 3 M$_\odot$.
For FG Ser, ER 
Del, and AG Peg, we 
have only lower limits for 
the bolometric magnitude. The range for 
V1261 Ori is indicated 
with the red dashed line. 
Adapted from Ref.~\citenum{Boffin14A}.}
   \end{figure}

\section{Symbiotic stars}
Symbiotic stars are a class of bright, variable red giant stars that show a composite spectrum, where on top of the typical absorption features of the cool star, there are strong hydrogen and helium emission lines, linked to the presence of a hot star and a nebula. These stars are interacting binary systems, with orbital periods between a hundred days and several years. The red giant in the system is loosing mass, part of which is transferred to the accreting companion -- either a main-sequence (MS) star or a white dwarf (WD).

We now know about 200 symbiotic stars, but have orbital elements for only 40 systems or so. There are, however, also similar systems to symbiotic stars, i.e. binaries with a red giant primary, but with lower mass-loss or mass transfer. Until very recently, it was not possible to firmly establish whether the mass-loss process in symbiotic and related stars took place via a wind or through Roche lobe overflow. Answering this question requires indeed to be able to compare the radius of the stars to the Roche lobe radius (which depends on the separation and the mass ratio). Several explanations have been proposed to account for this but the only way to answer it is by measuring the giant's radius of symbiotic stars in a direct way! Optical interferometry is currently the only available technique that can achieve this.

Using PIONIER, we therefore measured the diameters of several symbiotic and related stars. These diameters -- in the range of 0.6--2.3 mas -- are used to assess the filling factor of the Roche lobe of the mass-losing giants and provide indications on the nature of the ongoing mass transfer. We can also use this information to put the stars in an H-R diagram, as shown in Fig.~\ref{fig:diam}. Our analysis has shown that red giants in symbiotic systems are rather normal and obey similar relations between colour, spectral type, temperature, luminosity, and radius. Thus, the fact that they have larger mass-loss rates than single giants must be linked in some way to their binarity. 

For the three stars with the shortest orbital periods (i.e. HD 352, FG Ser and HD 190658), we find that the giants are filling (or are close to) their Roche lobe, consistent with the fact that these objects present ellipsoidal variations in their light curve. The other three studied stars (V1261 Ori, ER Del, and AG Peg) have filling factors in the range 0.2 to 0.55, i.e., the star is well within its Roche lobe. This led us\cite{Boffin14M} to suggest the existence of a dichotomy in symbiotic stars: systems which apparently fill their Roche lobes are those that contain a main-sequence companion or an helium (He) white dwarf and not a carbon/oxygen (CO) WD. It is, however, still difficult to understand why, if Roche lobe overflow takes place in our three systems, it does not lead to a dynamic common envelope, as the mass ratio we determine for our Roche-lobe filling giants are often larger than 1.5.  Clearly more theoretical work is needed along those lines.

 \begin{figure} [htbp]
   \begin{center}
   \includegraphics[width=15cm]{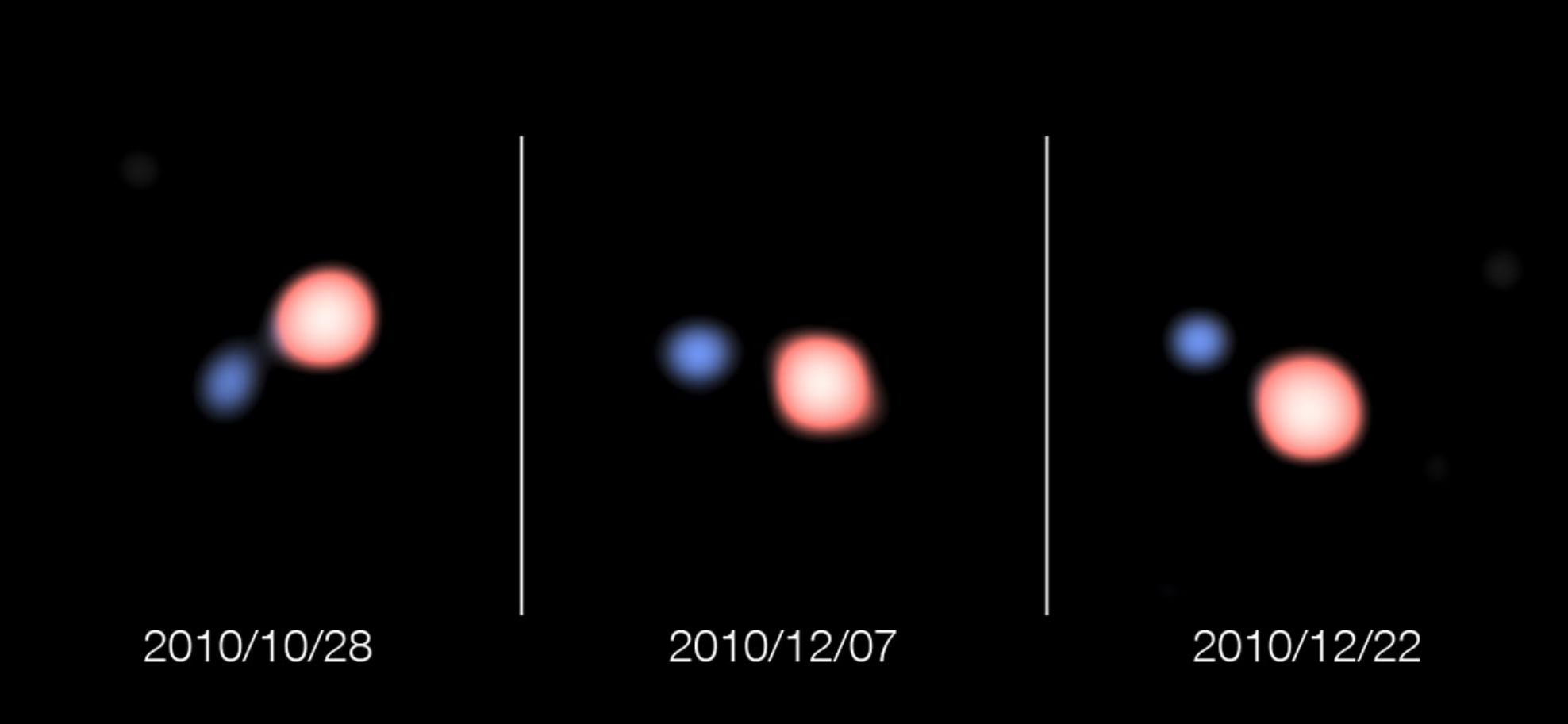}
   \end{center}
   \caption 
   { \label{fig:SSLep} 
Image reconstruction of our PIONIER observations of SS Lep showing the orbital motion of the resolved M giant and the A star. The distortion of the giant in the images is most certainly an artefact of the asymmetric PSF. Credit: ESO/PIONIER/IPAG; see Ref.~\citenum{Boffin14M}.}
   \end{figure} 
 
In one case, we have been able to study a symbiotic system in much more detail -- see Ref.~\citenum{Blind11}. The Algol system  
SS Leporis is composed of an evolved M giant and an A star in a 260-day orbit.

We were able to detect the two components of SS Lep as they moved across their orbit (see Fig.~\ref{fig:SSLep}) and measure the diameter of the red giant in the system ($\sim$2.2 mas). By reconstructing the visual orbit and combining it with the previous spectroscopic one, it was possible to well constrain the parameters of the two stars. The M giant is found to have a mass of 1.3 M$_\odot$, while the less evolved A star has a mass twice as large: a clear mass reversal must have taken place, with more than 0.7 M$_\odot$ having been transferred from one star to the other. Our results also indicate that the M giant only fills around 85$\pm$3\% of its Roche lobe, which means that the mass transfer in this system is most likely by a wind and not by Roche lobe overflow.

\begin{figure} [ht]
   \begin{center}
   \begin{tabular}{c} 
   \includegraphics[width=15cm]{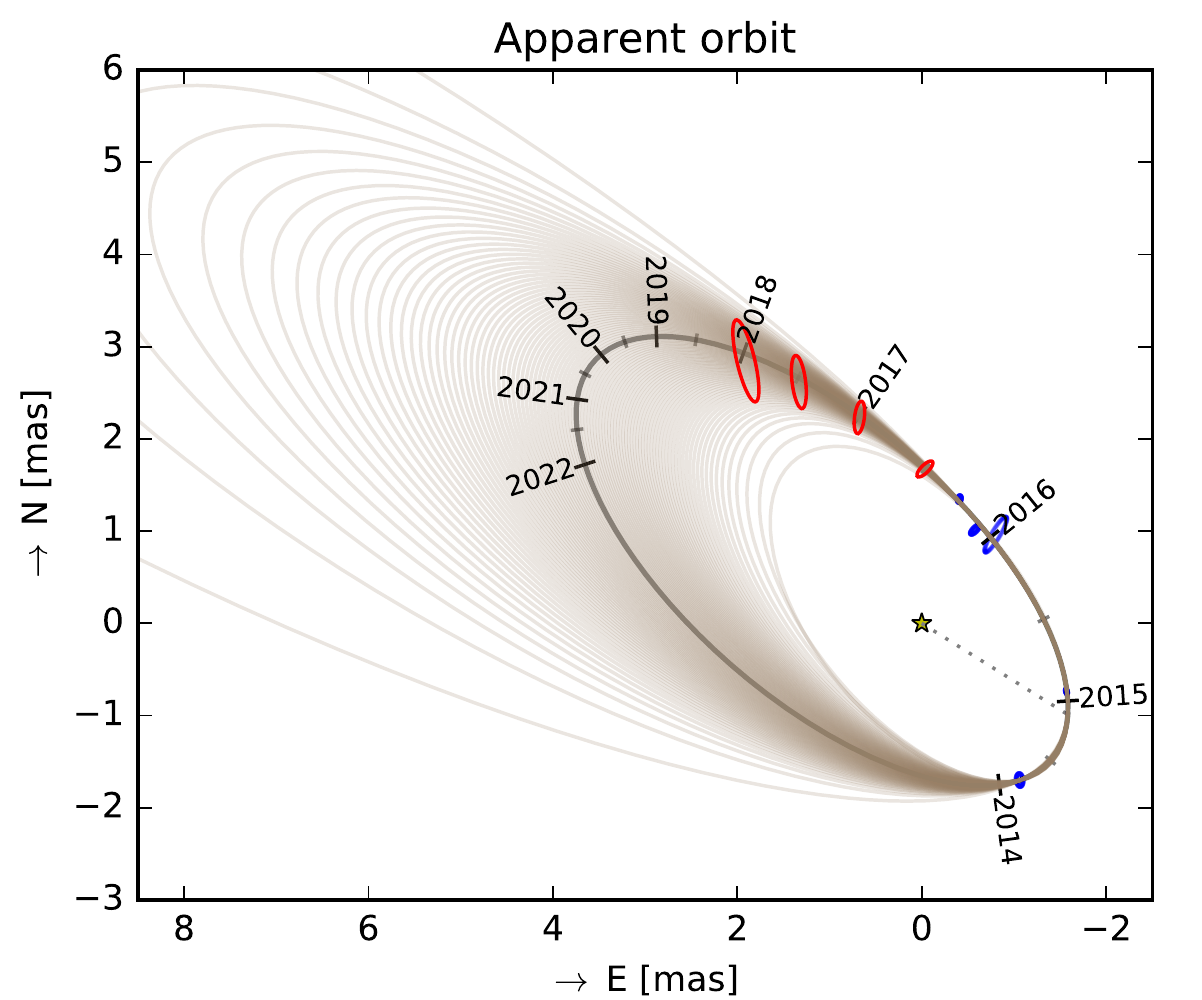}
   \end{tabular}
   \end{center}
   \caption 
   { \label{fig:hrcar} 
Family of orbits that fit all our PIONIER data points. The observed points are in blue, the projected orbits in grey and 1$\,\sigma$ ellipses allowing to distinguish between the various orbits in red. A large number of orbits still fit the data, with the longer orbits being the more eccentric. The right panel shows the de-projected orbits, in absolute scale, assuming a distance of 5.4 kpc. Adapted from Ref.~\citenum{Boffin16}.}
   \end{figure}

\section{Luminous Blue Variables}

Luminous Blue Variables (LBVs) are post-main-sequence massive stars undergoing extreme mass-loss and strong photometric and spectroscopic variability. The most extreme case, and best known, LBV, is {$\eta$\,Carinae}. 
 
There is currently a hot debate in the literature on the evolutionary status of LBV stars and on the importance of binarity in their formation. 
So far, however, while several wide LBV binaries were identified, LBV systems similar to $\eta$\, Car (with relatively short period and very eccentric) have not been found.  This is, at first glance, quite surprising as it is established than more than 70\% of all massive stars will exchange mass with a companion. However, LBVs are rare objects with complex emission line spectra and intricate nebulae, and they could also be the result of mergers, in which case they would {\it now} be single stars. Located at average distances of a few kilo parsecs or more, a direct search for a close companion requires at least milli-arcsecond spatial resolution -- only reachable by interferometry.

With PIONIER, we have been able to show that the LBV HR\,Car is in fact a binary system, with an orbital period of several years or several decades. 
Our interferometric observations clearly reveals its binary nature. We detected the orbital motion over a period of two years (Fig.~\ref{fig:hrcar}). It is still not possible to derive the orbital period which could be of the order of a few to several tens of years and the separation of the order of 10--270 au, but with the constraint that the largest orbit must also be the most eccentric, with a periastron distance most likely fixed around 2 mas, or 11 au. If the eccentricity is small and the orbit turns out to be of the order of 5 to 10 years, {\bf HR\,Car would be the second binary LBV presenting all the hallmarks and properties which make $\eta$~Car truly such a unique object}, but with components of much smaller masses. 

Apart from highlighting the possible role of binarity in the formation and/or evolutions of LBVs, the fact that HR Car is a binary is essential as it will allow us to derive the masses of the stars, which will be very useful to  compare to stellar evolutionary models. For now, we constrain the most likely range of total masses to be 33.5--45~M$_\odot$. 

\section{Going to the Visible}
We are now briefly considering what can be done if we have an interferometer working in the visible.
In Tab.~\ref{tab:distance}, we show the 
 absolute magnitude in the $V$ and $H$ bands for stars on the main-sequence and on the giant branch, as well as distances we can reach assuming that with PIONIER we can currently study objects with about $H$=8, while it is hoped that one can reach $V$=10 in the visible. 
 
\begin{table}[htbp]

   \caption{Distances reached in $V$ and $H$}
    \begin{center} { \begin{tabular}{@{} lrrrr @{}} 
      \hline
       Sp. Type & M$_V$ & M$_H$ & D$_V$ (pc) & D$_H$ (pc)\\
    \hline
O V&--5&-4&10,000 &2,500\\
B V&--3&--2.9&4,000&1,500\\
A V&1.5&1.3&500&215\\
G V&5&3.5&100&80\\
M V&9&4.8&16&44\\
WD&12&11.8&4&1.7\\
K III&0.5&--2.1&800&1000\\
M IIII&--0.6&--4.8&1300&3700\\
      \hline
   \end{tabular}}
   \end{center}
   \label{tab:distance}
\end{table}

It is clear from this table that going to the visible, with such limiting magnitudes, would allow us to reach more distant objects of early spectral types than currently possible, a similar number of objects for early giants (K-type), while slightly less of the reddest objects, that is M stars. However, one need to convolve this with the fact, that in the visible, the flux ratio
between a M star and its companion would be lower (the M star being redder), and it would thus be easier to detect the companion. This is especially interesting for M giants, such as those found in symbiotic stars (see above for one example).
It is also noteworthy that these numbers indicate that for O stars,  one can thus reach the Bulge and thus have access to many objects. 

If we could improve on current interferometers, and achieve a resolution of 0.1 mas, then a system at 10 kpc with such a separation, would imply a physical separation of 1 au. This is an interesting separation range to probe, as it is exactly those of interest for, e.g., LBVs. 

More generally, one can estimate for the different spectral classes, the range of orbital periods and mass ratios we could detect, if we assume we can observe objects up to $V$=10 with a resolution of 0.1 mas: \\
\pagebreak

\begin{table}[h]
\centering
\begin{tabular}{lll}
\multicolumn{3}{l}{\bf O V stars}\\
D=150 pc		& P $\sim$ 6 -- 1000 days & q $>$ 0.04 \\
D=1 kpc 		& P $>$ 100 d & q $>$ 0.1 \\
D=10 kpc		& P $>$ 9 yrs & q $\sim$ 1\\
\multicolumn{3}{l}{\bf A V stars}\\
D=10 pc		& P $<$ 100 d & q $>$ 0.1\\
D=100 pc		& P  $\sim$ 1 d to 10 yrs & q $>$ 0.25\\
D=500 pc		& P  $\sim$ 8 d to 100 yrs & q $\sim$ 1\\
\multicolumn{3}{l}{\bf G V stars}\\
D=10 pc		& P $<$ 70d & q $> $0.2\\
D=50 pc		& P $< $2 yrs&  q $>$ 0.5\\
D=100 pc		& P $<$ 6 yrs & q $\sim$ 1\\
\end{tabular}
\end{table}

The above numbers indicate that one could probe, for example, at least 250 Algol systems,
leading to a dramatic improvement in the understanding of such systems. This would be similar for all kinds of stars.
It is thus clear that the range of parameters that can be probed would be amazingly wide, and would provide much constraints for stellar evolution and binary processes!

\acknowledgments 
 
It is a pleasure to acknowledge my co-authors on all the papers based on interferometry published in the last 5 years. Special thanks go to Nicolas Blind, Jean-Louis Halbwachs, Jean-Baptiste LeBouquin, and Thomas Rivinius. 


\begin{thebibliography}{}
\bibitem{Blind11} Blind, N., Boffin, H. M. J., et al., ``An incisive look at the symbiotic star SS Leporis. Milli-arcsecond imaging with PIONIER/VLTI'', {\it A\&A} {\bf 536}, A55 (2011)
\bibitem{Boffin14M} Boffin, H. M. J. et al., ``A PIONIER View on Mass-transferring Red Giants'', {\it ESO Messenger} {\bf 156}, 35 (2014)
\bibitem{Boffin14A} Boffin, H. M. J. et al., ``Roche-lobe filling factor of mass-transferring red giants: the PIONIER view'', {\it A\&A} {\bf 564}, A1 (2014)
\bibitem{Boffin16} Boffin, H. M. J., et al., ``The LBV HR Car is Married'', {\it A\&A}, subm. (2016)
\bibitem{Halbwachs14} Halbwachs, J.-L., et al., ``Masses of the components of SB2 binaries observed with Gaia -- I. Selection of the sample and mass ratios of 20 new SB2s discovered with Sophie'', {\it MNRAS} {\bf 445}, 2371 (2014)
\bibitem{Halbwachs15} Halbwachs, J.-L., Boffin, H. M. J., et al., ``Accurate stellar masses for SB2 components: Interferometric observations for Gaia validation'', SF2A-2015, 377 (2015)
\bibitem{Halbwachs16} Halbwachs, J.-L., Boffin, H. M. J., et al., ``Masses of the components of SB2s observed with Gaia -- II. Masses derived from PIONIER interferometric observations for Gaia validation'', {\it MNRAS} {\bf 455}, 3303 (2016)
\bibitem{Rivinius15} Rivinius, T., Boffin, H. M. J., et al., ``Binarity of the LBV HR Car'', in [{\it New Windows on Massive Stars}]  {\bf 307}, 295 (2015)
\end{thebibliography}
\bibliographystyle{spiebib} 

\end{document}